\documentclass[a4paper,aps,twocolumn,nofootinbib,nobibnotes,twoside]{revtex4-2}

\usepackage{mathrsfs}
\usepackage[scr=rsfs]{mathalfa}
 
\usepackage{array}
\usepackage{geometry}
\usepackage{amsmath}
\usepackage{amsthm}
\usepackage{bm}
\usepackage{bbm} 
\usepackage{amssymb}
\usepackage{xspace}
\usepackage{amscd}
\usepackage{dsfont}
\usepackage{t1enc}
\usepackage{scalefnt}
\usepackage[greek, english]{babel}
\usepackage[latin2]{inputenc}
\usepackage{teubner}
\usepackage{graphicx}
\usepackage{multirow}
\usepackage{dcolumn}
\usepackage{braket}
\usepackage{xifthen}
\usepackage{textcomp}
\usepackage{tikz}
\usepackage{tipa}
\usepackage{pifont}
\usepackage{cancel}
\usepackage{stackrel}
\usepackage{xcolor}
\usepackage{soul}
\usepackage{stmaryrd}
\usepackage{multibib}

\usetikzlibrary{matrix,arrows,decorations.markings,decorations.pathreplacing,petri,topaths}
\usepackage[stable,symbol]{footmisc}
\definecolor{col}{rgb}{0.25, 0.41, 0.88}
\usepackage[pdftex, pdftitle={}, pdfauthor={}, colorlinks=true, urlcolor=col, linkcolor=col, bookmarks=false, citecolor=col, raiselinks=false,hyperfootnotes=false]{hyperref}

\usepackage{actaneasy}
\usepackage[T1]{fontenc}

\begin{document}

\ch{Smeared Field Description of\\ Free Electromagnetic Field}
\aum{J.A. Przeszowski} 
\arm{Smeared Field Description of Free Electromagnetic Field}
\spt{J.A. Przeszowski}{Smeared Field Description of Free Electromagnetic Field}
\aut{J.A. Przeszowski$^*$}{Faculty of Physics, University of Bia\l{}ystok, Cio\l{}kowskiego 4L, 15-424 Bia\l{}ystok, Poland}
\pacc{Free electromagnetic fields, satisfying Maxwell's equations with no charges and electric currents, can be described by complex vector fields. In the standard formulation with fields sharply dependent on position and time, one obtains integrals that are mathematically ill-defined. This happens for the massless Pauli--Jordan function, which is used to describe the time evolution of fields and appears in the Poisson brackets for classical fields. This difficulty can be solved by introducing smeared fields as linear functionals with test functions. In this way, the massless Pauli--Jordan function becomes a~tempered distribution, allowing a mathematically rigorous analysis.}{Riemann--Silberstein vectors, Poisson brackets, Pauli--Jordan function, tempered distributions\vs*{22pt}}{j.przeszowski@uwb.edu.pl}

{\bf 1. Introduction }

The electric and magnetic fields in empty space can be expressed in terms of a pair of complex vector fields $\we{F}(x)=\we{E}(x){-}\ii\, c\, \we{B}(x)$ and $\we{F}^*(x)=\we{E}(x){+}\ii\, c\, \we{B}(x)$, where $x=(t,\we{x})$ and $c=(\mu_0\epsilon_0)^{-1/2}$~[1]. This can be extended to an arbitrary homogeneous and isotropic dielectric~[2]. These fields are called the Riemann--Silberstein (RS) vectors~[3] and can be analysed in various aspects in both classical and quantum physics. In~[3] Bia\l{}ynicki-Birula claims that it is a complex vector-function of space and time coordinates that adequately describes the quantum state of a single photon. It is also argued that it can be practical for describing the quantum states of excitation of a free electromagnetic field, the electromagnetic field acting on a medium, the vacuum excitation of virtual electron--positron pairs, and for comparing the photon with other quantum particles that have their wave functions. Also, the Schr\"{o}dinger equation for a photon and the Heisenberg uncertainty relations can be formulated in terms of the RS vectors~[4]. More mathematical aspects of this formalism are presented in~[5] and~[6]. An overview of many features of classical and quantum electromagnetic fields described by RS vectors can be found in~[7]. 
In this article, we will discuss other aspects of classical fields in empty space, with \m{particular} reference to objects defined by momentum integrals that do not converge, as this can lead to self-inconsistency.
\mbox{}\\[5pt]
{\bf 2. Poisson brackets and temporal evolution}
\mbox{}\\[5pt]
In a free space with no charges and currents, the RS vectors allow us to express Maxwell's equations in a compact form 
\beq  
\aquad\  \ \begin{aligned}
&\partial_t \we{F}(x)=\ii c\, \nabla \times \we{F}(x), \
 &\nabla &\cdot\we{F}(x)=0, \\[12pt]
&\partial_t \we{F}^*(x)=-\ii c\, \nabla\times \we{F}^*(x),\ 
 & \nabla&\cdot\we{F}^*(x)=0.\\
\end{aligned}
\eeq{}\tyl
\beq
\eeq{1}
The electromagnetic energy density can be written as a simple expression if we scale the RS vectors by a constant factor 
\beq 
\begin{aligned}
& \mathcal{H}(x) =\frac{\epsilon_0}{2}\we{E}^2(x)+\frac{1}{2\mu_0}\we{B}^2(x)=\we{V}^*(x)\cdot \we{V}(x),\\[6pt]
& \we{V}(x)=\sqrt{\frac{\epsilon_0}{2}} \, \we{F}^*(x),\\[6pt]
& \we{V}^*(x)=\sqrt{\frac{\epsilon_0}{2} }\,  \we{F}^*(x).\\
\end{aligned} \eeq{} \tyl\tylx  \beq\eeq{2}
\noindent This leads to the Hamiltonian that generates the equations of motion for $\we{V}(x)$ and $\we{V}^*(x)$ by means of the Poisson brackets (PB) for fields at equal time $t=t_0$,
\beq 
\aquad 
\begin{aligned}
& \big\{ V_i(t_0,\we{x}), V_j^*(t_0,\we{y})\big\}_{_{\rm PB}} {=}\ \ii\, c\, \epsilon_{ikj}\, \partial^x_k\, \delta^3(\we{x}-\we{y}),&\\[6pt]
 &\big\{ V_i^*(t_0,\we{x}), V_j(t_0,\we{y})\big\}_{_{\rm PB}}  {=} -\ii\, c\, \epsilon_{ikj}\, \partial^x_k\, \delta^3(\we{x}-\we{y}),&\\[6pt]
& \big\{ V_i(t,\we{x}), V_j(t_0,\we{y})\big\}_{_{\rm PB}}\hs*{-1mm} {=} \big\{ V_i^*(t,\we{x}), V_j^*(t_0,\we{y})\big\}_{_{\rm PB}}\hs*{-1mm}{=}{0,}&\\
\end{aligned} \eeq{}\tyla \beq\eeq{3}\\[1pt]
where Einstein's summation convention for repeated indices is applied. Moreover, the equations of motion can be solved by means of the\linebreak integral~[8]\tylx\beq 
\begin{aligned}
&\we{V}(t,\we{x})= \int\limits_{\mathbb{R}^3}\dd{}^3\we{y} \bigg[ \frac{\partial }{\partial t} \,
D(t{-}t_0,\we{x}{-}\we{y}) \we{V}(t_0,\we{y})\\ 
\end{aligned}
\eeq{}\tyla
\beq
\quad \ -\ii c\, D(t{-}t_0,\we{x}{-}\we{y})\nabla\times\we{V}(t_0,\we{y})\bigg], 
\eeq{4}
where the function $D(t,\we{x})$ should satisfy the following properties 
\beq 
\begin{aligned}
&\big( \partial_t^2 -c^2 \nabla^2\big)\, D(x)=0,\quad \quad D(0,\we{x})=0,\\[6pt]
&\partial_t D(t,\we{x})\big|_{t=0}=\delta^3(\we{x}).\\[2pt]
\end{aligned}
\eeq{}\tyl\tylx\beq\eeq{5}\\
\indent The general solution (4) can be used to transform the Poisson brackets for equal time (3) into the generalized Poisson brackets for arbitrary instants~[8]\beq \aquad \ \
\begin{aligned}
&\big\{ V_i(t,\we{x}), V_j^*(t_0,\we{y})\big\}_{_{\rm {PB}}} =\\[6pt]
&\quad\  \left(c^2\delta_{ik} \nabla^2{-}c^2\partial_i\partial_k{+}\ii c\,\epsilon_{ikj}\partial^x_k \partial_t\right)D(t{-}t_0,\we{x}{-}\we{y}),\\[6pt]
&\big\{ V_i(t,\we{x}), V_j(t_0,\we{y})\big\}_{_{\rm {PB}}}=0\\
\end{aligned}
\eeq{}\tyl  \beq\eeq{6}\\
\indent The standard definition of $D(t,\we{x})$, called the massless Pauli--Jordan function~[9], is given by the momentum integral 
\beq 
D(t,\we{x})=\ii\int\limits_{\mathbb{R}^3} \frac{\dd{}^3\we{k}}{2\omega\, (2\pi)^3} \big(\ee^{-\ii k\cdot x}-\ee^{\ii k\cdot x}\big),
\eeq{7}
($k\cdot x= \omega t-\we{k}\cdot\we{x}$; $\omega=c|\we{k}|=ck$), which is clearly ill-defined. However, it is generally argued that this divergent integral defines a distribution that can be written explicitly using the Dirac delta distribution accordingly\beq 
D(t,\we{x})=\frac{{\rm sgn}(t)}{2\pi c}\, \delta\big(c^2t^2-\we{x}^2\big).
\eeq{8}\\[2pt]
\indent Formally, the standard definition (7) is the inverse Fourier transform of a function that is not integrable but is only locally integrable in $\mathbb{R}^3$. For this type of function, the corresponding Fourier transform can be consistently defined and calculated using tempered distributions, as recently proposed in~[10]. One might expect that the appearance of such divergent momentum integrals as in formula (7) is a consequence of the use of fields with a sharp dependence on the position vector. Therefore, we can try a different approach, using smeared fields to check whether the new results are consistent with the previous ones. 
\mbox{}\\[5pt]
{\bf 3. Smeared Riemann--Silberstein vectors}
\mbox{}\\[5pt]
Smeared fields are linear functionals for test functions, which vanish rapidly at infinity, so one may define smeared fields as integrals with test functions of the Schwartz class $\mathbb{S}(\mathbb{R}^3)\in f:~\mathbb{R}^3\to\mathbb{R}$~[11] 
\beq 
\begin{aligned}
&V_i[t,f]:=\int\limits_{\mathbb{R}^3}\dd{}^3\we{x}\ V_i(t,\we{x})f(\we{x}),\\
&V_i^*[t,f]:=(V_i[t,f])^*=\int\limits_{\mathbb{R}^3}\dd{}^3\we{x}\ V_i^*(t,\we{x})f(\we{x}).\\
\end{aligned}
\eeq{}\tyl\beq \eeq{9}\\[2pt]
\indent Thus, all these integrals converge, and integration by parts can be easily performed without the boundary term at infinity. This smearing can easily be applied to relations for fields with a sharp dependence on the position vector. From Maxwell's equations (1) one obtains the relations for smeared RS vectors 
\beq 
\begin{aligned}
&\partial_t V_i[t,f]=-\ii V_k[t,u_{ik}f], & \quad V_i[t,\partial_i f]=0,\\[7pt]
&\partial_t V^*_i[t,f]=\ii V^*_k[t,u_{ik}f], & \quad V_i^*[t,\partial_i f]=0,\\
\end{aligned}
\eeq{}\tyla \beq\eeq{10}\\[2pt] 
where $u_{ik}:=c\,\epsilon_{{ijk}}\partial_j$. Next, smearing of non-vanishing Poisson brackets for equal time (3) gives Poisson brackets for smeared RS vectors at equal time $t=t_0$
\beq 
\begin{aligned}
 &\{ V_i[t_0,f], V_j^*[t_0,g]\}_{_{\rm PB}}=\ii (u_{ij}g,f),\\[7pt]
&\{ V_i^*[t_0,g], V_j[t_0,f]\}_{_{\rm PB}}=-\ii (g,u_{ij}f),\\
\end{aligned}
\eeq{}\tyl\beq\eeq{11}\\[2pt]
where the inner product in the space of the Schwartz test functions $f,g\in \mathcal{S}(\mathbb{R}^3)$ is~[11]  
\beq 
(g,f)=\int\nolimits_{\mathbb{R}^3}\dd{}^3x\ g(x) f(x). 
\eeq{12} \\[1pt]
\indent The equations of motion for smeared fields can be easily diagonalized as 
\beq 
\begin{aligned}
&\partial_t a_i^{(+)}[t,f]=\ii a_i^{(+)}[t,\omega f],\\[6pt]
&\partial_t a_i^{(-)}[t,f]=-\ii a_i^{(-)}[t,\omega f],
\end{aligned}
\eeq{}\tyl\beq\eeq{13}\\[2pt]
with auxiliary smeared fields defined as
\beq 
\begin{aligned}
 a_i^{(+)}[t,f]:=V_i[t,\omega f]+V_j[t,u_{ij}f],\\[6pt]
 a_i^{(-)}[t,f]:=V_i[t,\omega f]-V_j[t,u_{ij}f],\\
\end{aligned}
\eeq{}\tyl\beq\eeq{14}\\[2pt]
where a new test function $\omega^a f:~\mathbb{R}^3\to\mathbb{R}$, is given by means of the inverse Fourier transform ($a\in \mathbb{N}$) 
\beq 
 (\omega^a f)(x):=\int\limits_{\mathbb{R}^3} \frac{\dd{}^3\we{k}}{(2\pi)^3}\, \ee^{\ii \we{k}\cdot \we{x}} (c|\we{k}|)^a \mathcal{F}\{f\}(k). 
\eeq{}\tyl\tylx\beq\eeq{15}
\indent The equations of motion for auxiliary smeared fields can be easily solved as 
\beq 
\begin{aligned}
&a_i^{(+)}[t,f]:= a_i^{(+)}[t_0,\ee^{\ii\omega (t-t_0)} f],\\[6pt]
&a_i^{(-)}[t,f]:= a_i^{(-)}[t_0,\ee^{-\ii\omega (t-t_0)} f],\\ 
\end{aligned}
\eeq{}\tyl \beq\eeq{16}\tyla 
\mbox{}
\noindent where one has 
\beq 
\big(\ee^{\pm \ii \omega t}f\big)(\we{x})=\sum^{\infty}_{n=0}\frac{(\pm \ii t)^n}{n!}\, (\omega^n f)(\we{x})=\\
\eeq{}\tylx \beq 
\quad \int\limits_{\mathbb{R}^3} \frac{\dd{}^3\we{k}}{(2\pi)^3}\ \ee^{\pm \ii\omega(\we{k})t}\mathcal{F}\{f\}(\we{k}).
\eeq{17}\\[3pt]
\mbox{}
These solutions enable us to write the smeared RS vector at an arbitrary instant $t$ as
\beq
\aquad \ 
\begin{aligned} 
&V_i[t,f]:=\frac{1}{2} \Big(a_i^{(+)}[t,\omega^{-1}f]+a_i^{(-)}[t,\omega^{-1}f]\Big)=\\[10pt]
&\quad V_i\big[t_0,\cos(\omega\tau)f\big]+\ii V_k\big[t_0, u_{ik} \sin(\omega \tau)\omega^{-1}f\big]=\\[10pt]
&\quad \int\limits_{\mathbb{R}^3}\dd{}^3\we{y}\ {\big(\cos(\omega\tau)f\big)(\we{y})}\, V_i(t_0,\we{y})\\[7pt]
\end{aligned}\eeq{}\tyla
\beq 
 -\ii \int\limits_{\mathbb{R}^3}\dd{}^3\we{y}\  {\big(\sin(\omega\tau)\,\omega^{-1}f\big)(\we{y})} \, U_i(t_0,\we{y}),
\eeq{18}
where $\tau=t-t_0$, $U_i(t_0,\we{x})=u_{ik}V_{k}(t_0,\we{x})$, and we extend the previous definition (15) to the case $a=-1$. When we integrate both sides of (4) with the test function $f(\we{x})\in \mathcal{S}(\mathbb{R}^3)$, assuming local integrability of $D(\tau,x)$, we obtain the tempered distribution $\mathcal{S'}(\mathbb{R}^3)$. Next, by switching the order of the integrals, we get a functional, which can be compared with~(18) 
\beq 
\aquad \ \begin{aligned}
&V_{i}[t,f]= \\ 
&\quad \int\limits_{\mathbb{R}^3}\dd{}^3\we{y}\ \frac{\partial}{\partial t}\left[ \ \int\limits_{\mathbb{R}^3}\dd{}^3\we{x}\ f(\we{x}) D(\tau,\we{x}{-}\we{y}) \right] V_i(t_0,\we{y})\\
&\quad -\ii  \int\limits_{\mathbb{R}^3}\dd{}^3\we{y}\ \left[\  \int\limits_{\mathbb{R}^3}\dd{}^3\we{x}\ f(\we{x}) D(\tau,\we{x}{-}\we{y})\right] U_i(t_0,\we{y}).
\end{aligned}
\eeq{}\tyl \beq\eeq{19}\\[3pt]
\indent Hence, we conclude that the massless Pauli--Jordan function $D(\tau,\we{x}{-}\we{y})$ satisfies the integral equation
\beq 
\begin{aligned}
\int\limits_{\mathbb{R}^3}\dd{}^3\we{x}\ f(\we{x})\, D(\tau,\we{x}{-}\we{y})=\left(\sin(\omega\tau)\omega^{-1}f\right)(\we{y}).
\end{aligned}
\eeq{}\tyl \beq\eeq{20}\\[2pt]
If we choose $t_0=0$ and $\we{y}=0$, then the integral equation (20) becomes the definition of the tempered distribution $D[t,f]$ as a linear functional 
\beq 
\aquad \ D[t,f]:=\hs*{-1mm}\int\limits_{\mathbb{R}^3}\hs*{-1mm}\dd{}^3\we{x}\ f(\we{x})\, D(t,\we{x})=\left(\sin(\omega t) \omega^{-1}f\right){(\we{0})}, 
\eeq{}\tyl\beq\eeq{21}\\[0pt]
which can be taken as a starting point for further analysis.   
\mbox{}\\[5pt]
{\bf 4. Analysis of the Pauli--Jordan functional~$D[t,f]$ }
\mbox{}\\[5pt]
Our analysis of $D[t,f]$ will use the calculation method proposed in~[10], thus we start with  
\beq 
\hs*{-3mm}\begin{aligned}
&D[t,f]=\left(\sin(\omega t)\omega^{-1}f\right)(\we{0})= \\[6pt]
&\quad \int\limits_{\mathbb{R}^3}\frac{\dd{}^3\we{k}}{(2\pi)^3}\ \frac{\sin(c k t)}{c k} \mathcal{F}\{f\}(\we{k})= \\[2pt]
\end{aligned}
\eeq{}\tyla 
\beq\quad
\hs*{-2mm} \int\limits_{\mathbb{R}^3}\frac{\dd{}^3\we{k}}{(2\pi)^3}\ \frac{\sin(c k t)}{c k} \left[\ \int\limits_{\mathbb{R}^3}\dd{}^3\we{x}\ f(\we{x})\ee^{-\ii \we{k}\cdot\we{x}}    \right].
\eeq{22}\\[3pt]
The next step requires switching the order of the integrals and if we do this directly in its present form, we get a divergent momentum integral, so such a final step would be mathematically incorrect,\vs*{-2pt}
\beq 
\aquad \ D[t,f]= 
\int\limits_{\mathbb{R}^3} \dd{}^3\we{x}\ f(\we{x}) \left[\ \int\limits_{\mathbb{R}^3} \frac{\dd{}^3\we{k}}{(2\pi)^3} \, \frac{\sin(c k t)}{c k} \ee^{-\ii \we{k}\cdot \we{x}} \right].
\eeq{}\tyla\beq\eeq{23}\\[2pt]
Note, however, that the divergent integral in square bracket in (23) is $D(t,\we{x})$, which was defined earlier by~(7). Since here it appears as the result of erroneous mathematical operations, hence we can conclude that the standard definition of~(7) is flawed or at best symbolic. For the smeared vector RS, we can avoid this pitfall, but we must carefully follow the steps below. 
\mbox{}
\indent Firstly, the Fourier transform of the test function in $\mathcal{S}(\mathbb{R}^3)$ allows integration by parts without a~boundary term, so we can perform the following transformation of integrals 
\beq \aquad \ 
D[t,f]=\int\limits_{\mathbb{R}^3}\hs*{-1mm}\frac{\dd{}^3\we{k}}{(2\pi)^3}\, \frac{\sin(c k t)}{c k}\left[\ \int\limits_{\mathbb{R}^3}\hs*{-1mm}\dd{}^3\we{x}\, f(\we{x})\ee^{-\ii \we{k}\cdot\we{x} }\right]\hs*{-1mm}{=}
\eeq{}
\beq
\int\limits_{\mathbb{R}^3}\hs*{-1mm}\frac{\dd{}^3\we{k}}{(2\pi)^3}\, \frac{\sin(c k t)}{c k}\frac{1}{k^2}\hs*{-1mm}\left[\ \int\limits_{\mathbb{R}^3}\hs*{-1mm}\dd{}^3\we{x}\, ({-}\Deltap ) f(\we{x})\ee^{-\ii\we{k}\cdot\we{x} }\right]\hs*{-1mm}{=}
\eeq{}
\beq
\int\limits_{\mathbb{R}^3}\hs*{-1mm}\dd{}^3\we{x}\, ({-}\Deltap )f(\we{x}) \int\limits_{\mathbb{R}^3}\hs*{-1mm}\frac{\dd{}^3\we{k}}{(2\pi)^3} \frac{\sin(c k t)}{c k^3}\ee^{-\ii \we{k}\cdot\we{x}}.
\eeq{24}\\[3pt]
This yields a convergent momentum integral, which can be calculated analytically using formula (3.741.3) in~[12],\beq 
\int\limits_{\mathbb{R}^3}\hs*{-1mm}\frac{\dd{}^3\we{k}}{(2\pi)^3}\ \frac{\sin(c k t)}{c k^3} \ee^{\ii \we{k}\cdot\we{x}} =
\eeq{}\tylx
\beq 
\quad \frac{{\rm sgn}(t)}{2\pi^2 c\, r}\int\limits_{\mathbb{R}^3}\dd{}k\ \frac{\sin(c\,k\, |t|)}{k^2} \sin(k\, r)=
\eeq{} \tylx 
\beq 
\quad \frac{{\rm sgn}(t)}{4\pi c} \left(1+\frac{c|t|{-}r}{r}\, \Thetap(r{-}c|t|)\right),
\eeq{25}\\[4pt]
where $r=|\we{x}|$ and $\mathbb{R}_{+}=\{x\in \mathbb{R}:~x\ge 0\}$. This leads to the final stage of the calculation, where we have to perform integration by parts for the convergent integral in $\mathbb{R}^3$. Omitting details, which will be presented elsewhere, we give the final result\beq 
D[t,f]=\frac{t}{4\pi} \int\limits_{\Omegap_3} \dd{}\omega_{\we{x}}\ f(c|t|\we{\hat{x}}), 
\eeq{26} \tylx 
\mbox{}
\noindent where $\dd{}\omega_{\we{x}}$ is the hypersurface element on the unit sphere $\Omega_3$ embedded in $\mathbb{R}^3$, with its surface area $|\Omega_3|=4\pi$ and $\we{\hat{x}}=\we{x}/r$ being the versor of the position vector. This formula is the main result of this work and can be used to study various properties of the massless Pauli--Jordan function $D(t,\we{x})$ for arbitrary time~$t$.
\mbox{}
First, one finds $D[0,f]=0\Longrightarrow D(0,\we{x})=0$ and
\beq
\frac{\dd{}}{\dd{}t}D[t,f]\Big|_{t=0}\hs*{-2mm}=\frac{1}{4\pi} \int\limits_{\Omegap_3} \dd{}\omega_{\we{x}}\, f(\we{0})=f(\we{0}) =
\eeq{}\tyla
\beq\quad
\int\limits_{\mathbb{R}^3} \dd{}^3\we{x}\ f(\we{x}) \delta^3(\we{x})\Longrightarrow \frac{\partial}{\partial t} D(t,\we{x})\Big|_{t=0}\hs*{-2mm}=\delta^3(\we{x}).
\eeq{}\tyl \beq\eeq{27}
\indent The other equations for $D[t,f]$ require more complicated calculations, but the use of equations presented in the Appendix can be quite helpful. First, using (39) in the Appendix, we obtain 
\beq 
\frac{\partial }{\partial t}D[t,f] =D[t,\partial_r(r\,f)]\Longrightarrow
\eeq{}\tyla
\beq\quad
\Longrightarrow \left(t\frac{\partial}{\partial t}+r\frac{\partial}{\partial t}+2\right) D(t,\we{x})=0,
\eeq{28}
which is the manifestation of the covariance at an infinitesimal dilation transformation. Then, (38) and (40) in the Appendix allow us to find 
\beq 
\frac{1}{c} \frac{\dd{}}{\dd{}t} D[t,x^i\, f]= c\,t\, D[t,\partial_i f]\Longrightarrow
\eeq{}\tyla
\beq\quad
\Longrightarrow \left( \frac{x^i}{c}\frac{\partial }{\partial t}+c\, t\frac{\partial }{\partial x^i}\right) D(t,\we{x})=0,
\eeq{29}
which is the manifestation of the invariance at an infinitesimal Lorentz boost transformation. Finally (41) and (42) in the Appendix lead to the d'Alambert equation of motion 
\beq 
\frac{\dd{}^2}{\dd{}t^2} D[t,f]=D[t,\nabla^2 f]\Longrightarrow
\eeq{}\tyla
\beq\quad
\Longrightarrow  \left(\frac{1}{c^2}\,\frac{\partial^2}{\partial t^2}-\nabla^2\right) D(t,\we{x})=0.
\eeq{30}
All the above implications are valid in the sense of the distributions $\mathcal{S'}(\mathbb{R}^3)$.
\mbox{}
Finally, we can give the explicit form of the distribution $D(t,\we{x})$ starting from the functional~(26), for which the Dirac delta distribution can be \m{introduced} according to the equations 
\beq
f(|a|)=\int\limits_{\mathbb{R}_{+}}\dd{}r\ \delta(r-|a|)f(r)=
\eeq{}\tyla 
\beq \quad 2\int\limits_{\mathbb{R}_{+}}\dd{}r\ r\,\delta(r^2-a^2) f(r),\eeq{31}
which are valid for $a\ne 0$. Thus for $t\ne 0$, we find two equivalent functionals 
\beq \aquad \ 
\begin{aligned}
& \int\limits_{\Omegap_3} \dd{}\omega_{\we{x}}\ f(c|t|\we{\hat{x}})= 
\int\limits_{\Omegap_3} \dd{}\omega_{\we{x}}  \int\limits_{\mathbb{R}_{+}} \dd{}r\ \delta(r{-}c|t|)\, f(r\we{\hat{x}})=\\
\end{aligned}
\eeq{} 
\beq 
\quad  \int\limits_{\mathbb{R}^3}\dd{}^3\we{x}\ \frac{\delta(r-c|t|)}{r^2} f(\we{x}),
\eeq{32}
\beq \aquad \ 
\begin{aligned}
& \int\limits_{\Omegap_3} \dd{}\omega_{\we{x}}\ f(c|t|\we{\hat{x}}){=} 
 2 \int\limits_{\Omegap_3} \dd{}\omega_{\we{x}}  \int\limits_{\mathbb{R}_{+}} \dd{}r\, \delta(r^2{-}c^2t^2)\, f(\we{\hat{x}})=\\
\end{aligned}
\eeq{}
\beq
\quad  2\int\limits_{\mathbb{R}^3} \dd{}^3\we{x}\ \frac{\delta(r^2-c^2t^2)}{r} f(\we{x}).
\eeq{33}\\[4pt]
\mbox{}
This leads to two equivalent expressions for 
\beq 
D(t,\we{x})=\frac{t}{4\pi}\frac{\delta(r{-}c\,|t|)}{r^2}=\frac{t}{2\pi}\frac{\delta(r^2{-}c^2t^2)}{r},
\eeq{34}\\[3pt]
and we can check that they satisfy the differential equations (28), (29), and (30), in the sense of distributions $\mathcal{S'}(\mathbb{R}^3)$. We must be aware that as long as $t\ne0$ these distributions are well defined, but for $t=0$ they would contain either $\delta(r)$ or $\delta(r^2)$ that are not well-defined distributions on $\mathbb{R}_{+}$. This caveat applies equally to formula~(7), which agrees with the second expression in~(34). Unfortunately, this caveat is usually omitted or even unknown, and therefore there are attempts to calculate $\partial_t D(t,\we{x})$ exactly at $t=0$, as in~[13], which cannot lead to the correct result. Moreover, the \m{Poisson} brackets for the sharp RS vectors at different instants of time, given by~(6), do not have a simple limit for equal times if we use (7) for the distribution, which implies the appearance of inconsistency. On the contrary, if we take the Poisson bracket for the smeared RS vectors, then from (11) and (18) we obtain the relation that is smooth at the\linebreak limit $\tau\to 0$, i.e.,
\begin{widetext} 
\beq 
\Big\{V_i^*[t_0,g], V_j[t,f]\Big\}_{_{\rm PB}}=-\ii \big(g,u_{ij}\cos(\omega\tau) f\big)+\big(g,u_{ik}\,u_{kj}\, \sin(\omega\tau)\,\omega^{-1}f\big)=
\eeq{}
\beq 
\quad -\ii \int\limits_{\mathbb{R}^3}\dd^3\we{x}\ g(\we{x})\, u_{ij}\, \big(\cos(\omega\tau)f\big)(\we{x})+
\int\limits_{\mathbb{R}^3}\dd^3\we{x}\  u_{ik}\,u_{kj}\, \big(\sin(\omega\tau)\, \omega^{-1}f\big)(\we{x}).
\eeq{35}\vs*{5pt}
\end{widetext} 
\mbox{}\\[5pt]
{\bf 5. Conclusions}
\mbox{}\\[5pt]
The smeared RS vectors correctly describe classical free electromagnetic fields, with no ill-defined mathematical expressions at any stage of the \m{calculations;} instead, tempered distributions appear naturally. We have explicitly calculated the distributions appearing in Poisson brackets and in the time evolution formula. Such analysis can be {extended} to both quantum electromagnetic field \m{theory} and massive fields. In particular, for \m{massive} fields we can determine the Pauli--Jordan \m{function} as a tempered distribution using an improved scheme to the one presented in~[13]. While the massless part carries the most singular contribution, the remaining part can be calculated quite easily. \vs*{5pt}  
\mbox{}\\[5pt]
{\bf Acknowledgments}
\mbox{}\\[5pt]
I would like to express my deep gratitude and respect to Professor Iwo Bia\l{}ynicki-Birula for the help and support he gave me during the preparation of my PhD thesis, suggesting its topic and being its kind and helpful supervisor.\vs*{5pt}
\mbox{}\\[5pt]
{\bf Appendix: Some useful integral equations}
\mbox{}\\[5pt]
By performing direct integration over the unit sphere embedded in $\mathbb{R}^3$, the following integral relations can be proved 
\beq 
\int\limits_{\Omegap_3} \dd{}\omega_{\we{x}}\ \partial_i f(\we{x})=\frac{1}{r^2}\int\limits_{\Omegap_3} \dd{}\omega_{\we{x}}\ \frac{\partial}{\partial r}  \Big[r\, x^i f(r\we{\hat{x}})\Big],
\eeq{}\tyl \beq \eeq{36}  
\beq 
\int\limits_{\Omegap_3} \dd{}\omega_{\we{x}}\ \nabla^2 f(r \we{x})=\frac{1}{r}\int\limits_{\Omegap_3} \dd{}\omega_{\we{x}}\ \frac{\partial^2}{\partial r^2}  \Big[r\, f(r\we{\hat{x}})\Big].
\eeq{}\tyl \beq \eeq{37}
Then, (36) applied to (26) gives $D[t,\partial_i f]$
\beq 
D[t,\partial_i f]=\frac{1}{4\pi c^2 t} \int\limits_{\Omegap_3} \dd{}\omega_{\we{x}}\ \left[\frac{\partial}{\partial r} \big(r x^i f(r\we{\hat{x}})\big)\right]_{r=c|t|}. \eeq{}\tyla\beq  
\eeq{38}
The temporal derivative of (26) is 
\beq 
\frac{\dd{}}{\dd{}t}D[t,f]=\frac{1}{4\pi}\int\limits_{\Omegap_3}\dd{}\omega_{\we{x}}\ \bigg[\frac{\partial}{\partial r} \Big(r f(r\we{\hat{x}})\Big)\bigg]_{r=c|t|},
\eeq{}\tyla\beq \eeq{39} 
which for a test function $x^i f(\we{x})\in \mathcal{S}(\mathbb{R}^3)$ takes the form 
\beq 
\frac{\dd{}}{\dd{}t}D[t,x^i f]=\frac{1}{4\pi}\int\limits_{\Omegap_3}\dd{}\omega_{\we{x}}\ \bigg[\frac{\partial}{\partial r} \Big(r x^i f(r\we{\hat{x}})\Big)\bigg]_{r=c|t|}.
\eeq{}\tyla\beq \eeq{40}
The second order temporal derivative of (26) is 
\beq \aquad \
\frac{\partial^2 }{\partial t^2}D[t, f]=\frac{c\, {\rm sgn}(t)}{4\pi}\int\limits_{\Omegap_3}\hs*{-1mm} \dd{}\omega_{\we{x}}\ \bigg[\frac{\partial^2}{\partial r^2}\Big( r f(r\we{\hat{x}})\Big)\bigg]_{r=c|t|}\hs*{0mm} .
\eeq{}\tyla \beq\eeq{41}
If one inserts (37) into (26), then one finds
\beq 
D[t, \nabla^2 f]= \frac{t}{4\pi\, c|t|}\int\limits_{\Omegap_3}\hs*{-1mm}  \dd{}\omega_{\we{x}}\ \bigg[\frac{\partial^2}{\partial r^2}\Big( r f(r\we{\hat{x}})\Big)\bigg]_{r=c|t|}\hs*{-1mm} . 
\eeq{}\tyla \beq \eeq{42}
\mbox{}\\[5pt]
{\bf References}
\mbox{}\\[5pt]
[1] {L. Silberstein, \refdo{{\em Ann. Phys.} {\bf 327}, 579 (1907)}{}} \\[2pt]
[2] {L. Silberstein, \refdo{{\em Ann. Phys.} {\bf 329}, 783 (1907)}{}} \\[2pt]
[3] {I. Bia\l{}ynicki-Birula, \refdo{\em Photon wave function}{10.1016/S0079-6638(08)70316-0}, {\em Progress in Optics}, Vol. 36, Ed. E. Wolf, Elsevier, Amsterdam 1996} \\[2pt]
[4] {I. Bia\l{}ynicki-Birula, Z. Bia\l ynicka-Birula, \refdo{{\em Phys. Rev. Lett.} {\bf 108}, 140401 (2012)}{10.1103/PhysRevLett.108.140401}}  \\[2pt]
[5] {A. Aste, \refdo{{\em J. Geom. Symmetry Phys.} {\bf 28}, 47 (2012)}{10.7546/jgsp-28-2012-47-58}} \\[2pt]
[6] {S.A. KHan, \refdo{{\em Phys. Scripta} {\bf 71}, 440 (2005)}{10.1238/Physica.Regular.071a00440}} \\[2pt]
[7]{I. Bia\l{}ynicki-Birula, Z. Bia\l ynicka-Birula, \refdo{{\em J. Phys. A: Math. Theor.} {\bf 46 }, 053001 (2013)}{10.1088/1751-8113/46/5/053001}, Corrigendum \refdo{{\em J. Phys. A: Math. Theor.} {\bf 46}, 159501 (2013)}{10.1088/1751-8113/46/15/159501.}}  \\[2pt] 
[8] {I. Bia\l{}ynicki-Birula, Z. Bia\l ynicka-Birula, \refdo{\em Quantum Electrodynamics}{}, Pergamon, Oxford 1975}  \\[2pt]
[9] {P. Jordan, W. Pauli, \refdo{{\em Z. Physik.} {\bf 47}, 151 (1928)}{}}  \\[2pt]
[10] {J.A. Przeszowski, E. Dzimida-Chmielewska, J.L. Cie\'{s}li\'{n}ski, \refdo{{\em Symmetry} {\bf 14}, 241 (2022)}{10.3390/sym14020241}}  \\[2pt]
[11]{I.M. Gel'fand, G.E. Shilov,  \refdo{\em Generalized Functions: Properties and Operations}{}, Vol. 1, orginal by E. Saletan (1958), translated from Russian, Academic Press, New York 1964 p.~192}  \\[2pt]
[12] {I.S. Gradshteyn, I.M. Ryzhik, \refdo{\em Table of Integrals, Series, and Products}{} 6th Ed., Academic Press, San Diego 2000}  \\[2pt]
[13] {Daqing Liu, Furui Chen, Shuyue Chen, Ning Ma, \refdo{{\em Eur. J. Phys.} {\bf 41}, 035406 (2020)}{10.1088/1361-6404/ab7496}}
\end{document}